\documentclass[%
preprint,
showpacs,preprintnumbers,
 amsmath,amssymb,
]{revtex4-1}

\usepackage{graphicx}% Include figure files
\usepackage{dcolumn}% Align table columns on decimal point
\usepackage{bm}% bold math
\usepackage{hyperref}% add hypertext capabilities
\begin{document}

\newcommand{\intsum}{\sum \kern -16pt \int}
\newcommand{\intsumup}{\sum \kern -16pt \int_{\alpha'}^{\bf u'}}
\newcommand{\intsumuz}{\sum \kern -16pt \int_{\alpha''}^{\bf u''}}
\newcommand{\intsumut}{\sum \kern -16pt \int_{\alpha'''}^{\bf u'''}}
\newcommand{\intsumvp}{\sum \kern -16pt \int_{\beta'}^{\bf v'}}
\newcommand{\intsumvz}{\sum \kern -16pt \int_{\beta''}^{\bf v''}}
\newcommand{\ua} {\, {\bf u} \, \alpha \,}
\newcommand{\uap} {\, {\bf u'} \, \alpha' \,}
\newcommand{\uaz} {\, {\bf u''} \, \alpha'' \,}
\newcommand{\uat} {\, {\bf u'''} \, \alpha''' \,}
\newcommand{\vb} {\, {\bf v} \, \beta \,}
\newcommand{\vbp} {\, {\bf v'} \, \beta' \,}
\newcommand{\vbz} {\, {\bf v''} \, \beta'' \,}

%\preprint{APS/123-QED}

\title{Solutions of the bound state Faddeev-Yakubovsky equations\\ in three dimensions
by using NN and 3N potential models}

\author{M. R. Hadizadeh$^1$}
\email{hadizade@ift.unesp.br}

\author{Lauro Tomio$^1$,$^2$}
\email{tomio@ift.unesp.br}

\author{S. Bayegan$^3$}
\email{bayegan@khayam.ut.ac.ir}

\affiliation{$^1$Instituto de F\'{\i}sica Te\'{o}rica (IFT), Universidade Estadual Paulista (UNESP), Barra Funda, 01140-070, S\~{a}o Paulo, Brazil,}
\affiliation{$^2$Instituto de F\'{\i}sica, Universidade Federal Fluminense, 24210-346, Niter\'oi, RJ, Brazil,}
\affiliation{$^3$Department of Physics, University of Tehran, P.O.Box 14395-547, Tehran, Iran.}

\date{\today}% It is always \today, today,

\begin{abstract}
A recently developed three-dimensional approach (without partial-wave decomposition) is considered to investigate solutions of Faddeev-Yakubovsky integral equations in momentum space for three- and four-body bound states, with the inclusion of three-body forces. In the calculations of the binding energies, spin-dependent nucleon-nucleon (NN) potential models (named, S$_{3}$, MT-I/III, YS-type and P$_{5.5}$GL) are considered along with the scalar two-meson exchange three-body potential.
Good agreement of the presently reported results with the ones obtained by other techniques are obtained, demonstrating the advantage of an approach in which the  formalism is much more simplified and easy to manage for direct computation.
\end{abstract}

\maketitle

\section{Introduction}

In recent years, calculations of three- and four-body bound and scattering states based
on the Faddeev-Yakubovsky (FY) scheme are performed in a novel three-dimensional (3D)
approach, which avoids truncation problems and the necessity of complicated recoupling algebra that accompanies partial-wave (PW) based calculations \cite{Schellingerhout-PRC46}-\cite{Epelbaum-PRC70}. Instead, in the 3D approach, the equations and amplitudes are formulated directly as functions of momentum vector
variables. This is a straightforward procedure quite convenient for obtaining final
observables such as the total energy.
For a PW observable, one can easily project the final state onto the
specific required partial-wave channel.

For three-nucleon (3N) and four-nucleon (4N) bound states, the FY equations with two-
and three-nucleon interactions have been recently formulated in a realistic 3D
approach \cite{Bayegan-PRC77}.
The formalism, according to the number of spin-isospin states that one takes into account,
leads to finite number of coupled three dimensional integral equations to be solved. It has
been shown that considering the continuous angle variables instead of the discrete angular
momentum quantum numbers in evaluation of the transition and permutation operators, coordinate
transformations as well as the three-nucleon forces (3NFs) lead to less complicated expressions
in comparison with the PW representation. However, it should be mentioned that with respect to
the PW representation, the present formalism with the smaller number of equations leads to higher
dimensionality of integral equations. In other words the price for the smaller number of equations
in 3D representation is the higher dimensionality of the integral equations. It should be clear
that by switching off the spin-isospin quantum numbers, one can easily reach the bosonic type of
three dimensional FY integral equations which are solved in Refs. \cite{Elster-FBS27}-\cite{Hadizadeh-EPJA36}.

In view of the above, we can observe that one real advantage in using a non-PW
approach in comparison with PW-based methods relies in a simplified computational
algorithm, which is straightforward obtained from the original equations.
For interacting systems with two and three particles the procedure was already shown
to be quite reliable and easy to be implemented. The advantage of the 3D approach
is more evident in the formulation of 4N interacting systems, where it completely avoids
the extremely complicated algebra of coupling of spin-angular momentum quantum numbers.
However, it is clear that such advantage of the 3D approach, when dealing with the
formalism and the corresponding computation, comes at the expense of possible numerical
precision when considering more than two Jacobi momentum vector variables. In such a
case, by working with the non-PW approach, after the momentum variable discretization
one may have to deal with matrices larger than the ones that occur in case of PW-based
calculations, making the latter procedure preferable.

By considering previous numerical comparisons between 3D and PW-based results, we should
note the perfect agreement between the obtained full wave function of three-nucleon system, as well as the corresponding momentum distribution functions \cite{Elster-FBS27}.
In view of these results, in case of a four-nucleon interacting system, the numerical
accuracy obtained by the 3D approach is expected to be about the same as the accuracy
verified in PW-based calculations. This agreement should show up in the analysis of the
corresponding observables, which is partially done in the present approach by considering
bound-state solutions of three- and four-nucleon systems with 3NFs.

The 3D approach has been shown to be efficient in solving the Faddeev equations for the
3N scattering calculations, especially at intermediate and higher energies \cite{Liu-PRC72}.
Also, the recent proton-deuteron elastic and breakup calculations show that the 3D approach has the potential to provide a more rigorous treatment of Coulomb effects \cite{Witala-EPJA41}.

In the case of continuum problems, as for example when obtaining scattering
observables, where partial-wave summation can be problematic, the 3D approach
is expected to be particularly more efficient than a method using PW decomposition.
Clearly, intrinsic limitations of the PW-based calculations are not only due to the
complexity of deriving the necessary equations, but also due to the limitations
in computer resources requiring very large number of angular momentum states in order
to achieve convergence for the scattering observables.
By increasing the energy the number of PW channels strongly proliferates and consequently
leads to more numerical difficulties with respect to accuracy and storage requirements.
However, as shown in Ref. \cite{Liu-PRC72}, relativistic three-body scattering calculations
at energies up to 1 GeV laboratory kinetic energy has been done successfully by using
direct vector variable calculations, avoiding PW decomposition.
Since the 3D approach does not use partial wave decomposition, carrying all the PW
channels automatically, the same numerical effort is spent in observable calculations at
higher or lower energies.
Essentially, the 3D technique is not only shown to be a viable alternative to the
well-established PW-based calculations at low-energy regions, but also it appears to be a
necessary approach at higher energies where the PW approach is no longer feasible.

One should also note that channel independent observables, such as the
total differential cross section, can be obtained using the 3D formalism and consequently be compared to experimental
data. Since experimental data are not always available, one needs to extract from this
3D approach a channel-dependent observable, as the $NN$ phase shifts. To this aim,
one can easily project the obtained final state onto the specific PW channel, as it was
done by Fachruddin, leading to very accurate results in excellent agreement with established
PW results \cite{Fachruddin-PhD}.

Before concluding this introduction, it is useful to mention a recent alternative 3D
representation for 3N bound states where the spin-isospin couplings are not explicitly
carried out \cite{Gloeckle-FBS47}.
The novelty of this formalism is the evaluation of NN $t$-matrices, the 3NFs, and the
Faddeev components as products of scalar functions with scalar products of spin operators
and momentum vectors. The spin operators have been removed and the final formalism leads
to scalar functions of only momentum vectors.

In the present paper, our purpose is to calculate FY bound-state solutions
using nucleon-nucleon potential models with three-nucleon forces, following the non-PW 3D
approach as shown in Ref. \cite{Bayegan-PRC77}.
We report results obtained for three- and four-nucleon binding energies by employing
spin-isospin dependent NN potential models along with a scalar two-meson exchange 3NF.
The main goal of the present work is to demonstrate advantages of the 3D approach in
few-body systems by testing the 3D representation of the FY integral equations with
several potential models not previously considered in 3D approach studies.

The current paper is organized as follows: In section \ref{sec: FY}, we briefly review the coupled
three-dimensional FY integral equations for the $4N$ bound state. In section \ref{sec: Numerical results},
we present our numerical results for three- and four-nucleon binding energies and compare them
to the results obtained from other techniques. Finally, we have our summary with an outlook
in section \ref{sec: Summary}.

\section{A Brief Review of FY Equations in Three-Dimensions } \label{sec: FY}
\begin{figure}
\begin{center}
\includegraphics[width=10cm]{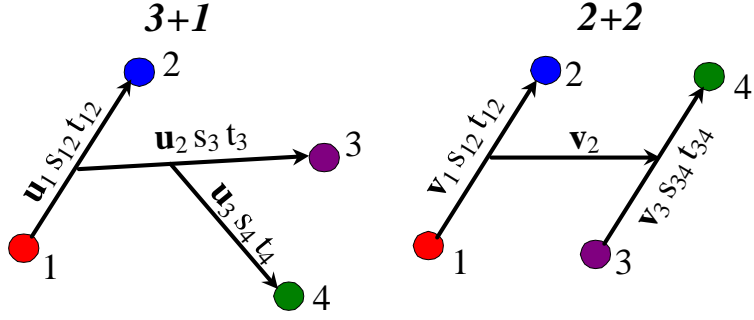}% Here is how to import EPS art
\end{center}
\caption{\label{fig:basis_states} Definition of the $3+1$ and $2+2$
type of Jacobi coordinates of a 4N system.}
\end{figure}

In the FY formalism, the bound state of four nucleons in the presence of 3NFs is described by the
following coupled equations \cite{Nogga-PRC65}:
\begin{eqnarray}
\label{eq.YCs} |\psi_{1}\rangle &=& G_{0}tP
\biggl [(1-P_{34})|\psi_{1}\rangle+|\psi_{2}\rangle \biggr] +(1+G_{0}t) G_{0}
V_{123}^{(3)} |\Psi\rangle, \nonumber \\ |\psi_{2}\rangle &=&
G_{0}t\tilde{P} \biggl[(1-P_{34})|\psi_{1}\rangle+|\psi_{2}\rangle \biggr],
\end{eqnarray}
where the Yakubovsky components $|\psi_{1}\rangle$ and $|\psi_{2}\rangle$ stand for ``3+1" (K-type or
{123,4}) and ``2+2" (H-type or {12,34}) partitions of the four nucleons, respectively.
$G_{0}$ is the free $4N$ propagator, the operator $t$ is the $NN$ transition matrix, and $P, P_{34}$
and $\tilde{P}$ are permutation operators. The quantity $V_{123}^{(3)}$ defines a part of the 3NF in
the cluster $(123)$,
which is symmetric under the exchange of particles $1$ and $2$.
As shown in Fig. \ref{fig:basis_states} for non-PW momentum space representation of the coupled Yakubovsky
components, i.e. Eq. (\ref{eq.YCs}), one needs two different sets of basis states:
\begin{eqnarray}
\label{eq.basis_states}
  |\, {\bf u} \,\,  \alpha  \, \rangle &\equiv&
  \Bigl | \, {\bf u}_{1}\,{\bf u}_{2}\,{\bf u}_{3} \,\,  \alpha^{S}_{1234} \, \alpha^{T}_{1234}\, \,  \Bigr \rangle
  \equiv  \Bigl | \, {\bf u}_{1}\,{\bf u}_{2}\,{\bf u}_{3} \,\,
 \Bigl ( \, (s_{12} \,\, \frac{1}{2})s_{123} \,\, \frac{1}{2} \, \Bigr) S \, M_{S}   \,\,
  \Bigl ( \, (t_{12} \,\, \frac{1}{2})t_{123} \,\, \frac{1}{2} \, \Bigr ) T \, M_{T} \, \,  \Bigr \rangle, \nonumber \\
 |\, {\bf v} \,\,  \beta  \, \rangle &\equiv&
 \Bigl |\, {\bf v}_{1}\,{\bf v}_{2}\,{\bf v}_{3} \,\, \beta^{S}_{1234} \, \beta^{T}_{1234} \, \Bigr \rangle
 \equiv \Bigl |\, {\bf v}_{1}\,{\bf v}_{2}\,{\bf v}_{3} \,\, (s_{12}\,\, s_{34}) S \, M_{S} \,\, (t_{12}\,\, t_{34}) T \, M_{T} \, \Bigr \rangle,
\end{eqnarray}
where these basis states are complete in the $4N$ Hilbert space:
\begin{eqnarray}
\label{complete}
 \intsum_{\xi}^{\bf A}  \,\,|\, {\bf A} \, \xi \,\rangle \,\langle\, {\bf A}\, \xi \,|=
\mathbf{1} , \,\, \intsum_{\xi}^{\bf A}\equiv\sum_{\xi} \int D^{3}A
\equiv\sum_{\xi} \int d^{3}A_{1}\, \int d^{3}A_{2}\, \int
d^{3}A_{3},
\end{eqnarray}
where ${\bf A}$ indicates each one of the ${\bf u}$ and ${\bf v}$ vector sets and $\xi$ indicates $\alpha$
and $\beta$ quantum number sets. Representation of the coupled equations (\ref{eq.YCs}) in the
introduced basis states, Eq.~(\ref{eq.basis_states}), leads to two sets of coupled integral equations:
\begin{eqnarray}
 \langle {\ua} |\psi_{1}\rangle &=& \intsumup \intsumuz   \langle {\ua}|G_{0}t | \uap \rangle \, \langle
\uap| P| \uaz \rangle\, \nonumber \\* &\times&   \left ( \, \intsumut  \, \langle \uaz |1-P_{34}| \uat \rangle
 \, \langle \uat |\psi_{1}\rangle  +  \intsumvp  \langle \uaz | \vbp \rangle \,\langle
 \vbp |\psi_{2}\rangle \, \right) \nonumber \\* &+&
\intsumup \intsumuz \langle \ua |(1+G_{0}t) G_{0} | \uap \rangle \,
\langle \uap | V_{123}^{(3)}| \uaz \rangle \langle \uaz |\Psi\rangle,
 \nonumber \\*   \\*
\langle \vb |\psi_{2}\rangle &=& \intsumvp \intsumvz \, \langle \vb |G_{0}t| \vbp \rangle \, \langle
\vbp | \tilde{P}| \vbz \rangle \, \nonumber
\\* &\times&  \left ( \, \intsumup \intsumuz \langle \vbz | \uap \rangle \, \langle \uap |1+P_{34}|
\uaz \rangle \, \langle \uaz | \psi_{1}\rangle +  \langle \vbz | \psi_{2}\rangle \, \right). \nonumber
\label{eq.YCs_basis}
\end{eqnarray}

To evaluate the above coupled integral equations, one needs to evaluate the matrix elements of two-body
$t$-matrices, permutation operators, as well as the coordinate transformations. These have been
evaluated in detail in Ref. \cite{Bayegan-PRC77}. It is useful to mention that one needs the free
4N basis states $|\, {\bf A} \,\, \gamma \, \rangle$, where the spin-isospin parts $\gamma$ are given
as: $|\gamma \,
\rangle \equiv| \gamma_{S} \,\, \gamma_{T} \,
 \rangle \equiv | m_{s_{1}} \, m_{s_{2}} \,  m_{s_{3}} \,  m_{s_{4}} \,\, m_{t_{1}} \, m_{t_{2}} \,
 m_{t_{3}} \, m_{t_{4}} \, \rangle.  $
In changing the 4N basis states, i.e. $| \alpha \, \rangle$ and $| \beta \, \rangle$, to
the free 4N basis states $| \gamma \, \rangle$, one needs to
calculate the usual Clebsch-Gordan coefficients
$\langle \, \gamma | \alpha \, \rangle = g_{\gamma \alpha } \equiv g_{\gamma \alpha }^{S} \, g_{\gamma \alpha }^{T}$ and
$\langle \, \gamma | \beta \, \rangle = g_{\gamma \beta } \equiv g_{\gamma \beta }^{S} \, g_{\gamma \beta }^{T}$ (see Ref. \cite{Bayegan-PRC77}).
After the mentioned operators and coordinate transformations are carried out, the coupled Yakubovsky equations
can be obtained explicitly:

\begin{eqnarray}
\label{eq.YCs_evaluated} \langle \, {\bf u}\, \alpha \, |\psi_{1}\rangle &=&
\frac{1}{{E-\frac{u_{1}^{2}}{m}
-\frac{3u_{2}^{2}}{4m}-\frac{2u_{3}^{2}}{3m}}}  \nonumber
\\* && \hspace{-20mm} \times \Biggl [\,
  \int d^{3}u_{2}' \, \sum_{\gamma',\gamma'''} \,
g_{\alpha \gamma'''} \, \delta_{m'''_{s_{4}} m'_{s_{4}}}
\delta_{m'''_{s_{3}} m'_{s_{1}}} \, \delta_{m'''_{t_{4}} m'_{t_{4}}}
\delta_{m'''_{t_{3}} m'_{t_{1}}} \nonumber
\\*  &\times& \,\, _{a} \Bigl \langle{\bf u}_{1}\, m'''_{s_{1}}
m'''_{s_{2}} \, m'''_{t_{1}} m'''_{t_{2}} \, \Bigl | \, t(\epsilon)
\, \Bigr | \, \frac{-1}{2}{\bf u}_{2} -{\bf u}'_{2} \, m'_{s_{2}} m'_{s_{3}} \,
m'_{t_{2}} m'_{t_{3}} \Bigr \rangle_{a} \, \nonumber
\\* &\times& \Biggl\{\,\, \sum_{\alpha''} \, g_{\gamma' \alpha''} \Bigl \langle{\bf u}_{2}+\frac{1}{2} {\bf u}'_{2}
\,\, {\bf u}'_{2}\,\,{\bf u}_{3} \, \alpha'' \, \Bigr | \psi_{1} \Bigr \rangle
\nonumber
\\* \quad && \hspace{2mm} - \sum_{\alpha''} \, g_{\gamma'_{1243} \alpha''} \, \Bigl \langle{\bf u}_{2}+\frac{1}{2}
{\bf u}'_{2} \,\, \frac{1}{3}{\bf u}'_{2}+ \frac{8}{9}{\bf u}_{3}
\,\, {\bf u}'_{2}-\frac{1}{3}{\bf u}_{3} \, \alpha''
\, \Bigr |\psi_{1} \Bigr \rangle
 \nonumber \\*  && \hspace{2mm} + \, \sum_{\beta'} \, g_{\gamma'
 \beta'} \,
 \Bigl \langle{\bf u}_{2}+\frac{1}{2} {\bf u}'_{2}\,\, -{\bf u}'_{2}-\frac{2}{3}{\bf u}_{3} \,\,
\frac{1}{2}{\bf u}'_{2}-\frac{2}{3}{\bf u}_{3} \, \beta'
\, \Bigr |\psi_{2} \Bigr \rangle\,\, \Biggr\} \nonumber \\*  && \hspace{-17mm} +
 \Biggl\{\,\, \Bigl \langle \, {\bf u}\, \alpha \, \Bigl |V_{123}^{(3)}
\, \Bigr |\Psi \Bigr \rangle \nonumber \\  && \hspace{-10mm} + \frac{1}{2}
\sum_{\gamma',\gamma'',\alpha'''} \,  g_{\alpha \gamma'} \,
g_{\gamma'' \alpha'''}  \int d^{3}u_{1}'\,
\frac{\delta_{m'_{s_{3}} m''_{s_{3}}} \delta_{m'_{s_{4}}
m''_{s_{4}}} \delta_{m'_{t_{3}} m''_{t_{3}}} \delta_{m'_{t_{4}}
m''_{t_{4}}} }{E-\frac{u_{1}'^{2}}{m}
-\frac{3u_{2}^{2}}{4m}-\frac{2u_{3}^{2}}{3m}} \nonumber \\* &&
\hspace{-10mm} \quad \times \, _{a} \Bigl \langle{\bf u}_{1}\, m'_{s_{1}}
m'_{s_{2}} \, m'_{t_{1}} m'_{t_{2}} \, \Bigl |t(\epsilon) \Bigr |{\bf u}'_{1} \,
m''_{s_{1}} m''_{s_{2}} \, m''_{t_{1}} m''_{t_{1}} \Bigr \rangle_{a}
\, \Bigl \langle{\bf u}'_{1}\,{\bf u}_{2}\,{\bf u}_{3} \,
\alpha'''\, \Bigl |V_{123}^{(3)} \, \Bigl |\Psi \Bigr \rangle \Biggr\} \, \, \Biggr],
 \nonumber \\*  \\*
\langle \, {\bf v}\, \beta \, |\psi_{2}\rangle &=&
\frac{1}{E-\frac{v_{1}^{2}}{m}
-\frac{v_{2}^{2}}{2m}-\frac{v_{3}^{2}}{m}} \nonumber \\* &\times&
\int d^{3}v_{3}' \, \, \sum_{\gamma',\gamma'''} \, g_{\beta
\gamma'''} \, \delta_{m'''_{s_{3}} m'_{s_{1}}} \,
\delta_{m'''_{s_{4}} m'_{s_{2}}} \, \delta_{m'''_{t_{3}}
m'_{t_{1}}} \, \delta_{m'''_{t_{4}} m'_{t_{2}}} \nonumber \\*
&\times& \,\, _{a} \Bigl \langle {\bf v}_{1} \, m'''_{s_{1}} m'''_{s_{2}}
\, m'''_{t_{1}} m'''_{t_{2}} \, \Bigl |t(\epsilon^{*}) \Bigr| {\bf v}'_{3}
 \, m'_{s_{3}} m'_{s_{4}} \, m'_{t_{3}} m'_{t_{4}} \Bigr \rangle _{a}\, \nonumber
\\* &\times& \Biggl \{\,\, \sum_{\alpha'} \, g_{\gamma' \alpha'} \, \Bigl \langle{\bf v}_{3}\,\,
 \frac{2}{3}{\bf v}_{2}+\frac{2}{3}{\bf v}'_{3} \,\, \frac{1}{2}{\bf v}_{2}-{\bf v}'_{3} \, \alpha' \, \Bigl |\psi_{1} \Bigr \rangle
  \nonumber \\* && \hspace{2mm} - \sum_{\alpha'} \, g_{\gamma'_{1243} \alpha'} \, \Bigl \langle{\bf v}_{3}\,\,
 \frac{2}{3}{\bf v}_{2}-\frac{2}{3}{\bf v}'_{3} \,\, \frac{1}{2}{\bf v}_{2}+{\bf v}'_{3} \, \alpha' \Bigr |\psi_{1} \Bigr \rangle
\nonumber \\*
  && \hspace{2mm}+ \sum_{\beta'} \, g_{\gamma' \beta'} \,
\Bigl \langle{\bf v}_{3}\,\,-{\bf v}_{2}\,\, {\bf v}'_{3} \, \beta'
\Bigr |\psi_{2} \Bigr \rangle \,\, \Biggr\}, \nonumber
 \end{eqnarray}
where $_{a}\langle \,|t(\epsilon)|\, \rangle _{a}$
and $_{a}\langle \,|t(\epsilon^{*})|\, \rangle _{a}$ are anti-symmetrized NN $t$-matrices.
This spin-isospin 3D formalism can be simplified to the bosonic case by switching off the
spin-isospin quantum numbers (see Refs. \cite{Hadizadeh-FBS40,Hadizadeh-EPJA36}).

\section{Numerical Results for the Three- and Four-Nucleon Binding Energies }\label{sec: Numerical results}
In this section, we present numerical results for the three- and four-nucleon binding energies.
The details of the numerical algorithm for solving the coupled three-dimensional integral equations can be
found in Refs. \cite{Bayegan-PRC77,Hadizadeh-FBS40,Hadizadeh-EPJA36}.

\subsection{Results for $NN$ Potential Models}

In order to check our proposed 3D formulation for the three- and four-nucleon bound states, we apply the
formalism to the following spin-dependent NN potential models: S$_{3}$
\cite{Afnan-PR175}, YS \cite{Gibson-PRC15}, MT-I/III
\cite{Malfliet-NPA127} and P$_{5.5}$GL \cite{Gibson-PRC18}. We are aware that realistic NN potentials have already been
used even for nuclei with $A>4$, but the main goal of the present work is the test of the 3D representation of
the FY equations for more realistic potentials that we have been used before in such 4B calculations.
The parameters of the above potentials are given in Table \ref{table.potentials}.

% potentials parameters
\begin{table}
\caption{\label{table.potentials}List of parameters of the NN potentials used
in this work. Each potential contains two parts, $V_{0}$ and $V_{1}$
where the indices 0 and 1 denote the spin of the $2N$ subsystem.
Each part is written as a sum of a few terms; each is expressed as
$V_{si} \, f(\mu_{si},r(p,p'))$, where
$f(\mu_{si},r)=exp(-\mu_{si}\,r^{2})$ for Gauss-type potential,
$f(\mu_{si},r)=exp(-\mu_{si}\,r)/r$ for Yukawa-type potential and
$f(\mu_{si},p,p')=\frac{ \xi_{i}^2}{m}. \,
\frac{(p\,p')^{2i-2}}{(p^{2}+\mu_{si}^{2})^{i} (p'^{2}+\mu_{si}^{2})^{i}}$ for separable potentials.
The potential strengths $V_{si}$ are in MeV for S$_{3}$, in
$fm^{-3}$ for YS and $P_{5.5}GL$ and dimensionless for MT-I/III. The
range parameters, exchanged masses for MTI/III, $\mu_{si}$ are in
$fm^{-2}$ for S$_{3}$ and in $fm^{-1}$ for others. For separable
potentials $\xi_{1}=1.0000$ and $\xi_{2}=2.9499$.}
\begin{center}
\begin{tabular}{lccclccccccccccccccccccccccccc}
\hline\hline Potential &&&&   Type       &&&&  i    &&&&  $V_{0i}$ &&&&
$\mu_{0i}$ &&&& $V_{1i}$ &&&& $\mu_{1i}$  \\ \hline
S$_{3}$   &&&&   Gauss      &&&&  1    &&&&  1000.0   &&&&     3.00   &&&&   1000.0 &&&& 3.00       \\
          &&&&              &&&&  2    &&&&  -326.7   &&&&     1.05   &&&&   -166.0 &&&& 0.80        \\
          &&&&              &&&&  3    &&&&  43.0     &&&&     0.60   &&&&    23.0  &&&& 0.40       \\
\\
YS &&&&   Separable \\
YS-I      &&&&              &&&& 1     &&&&  -0.1490   &&&& 1.165      &&&&  -0.4160  &&&&1.450  \\
YS-II     &&&&              &&&& 1     &&&&  -0.1430   &&&& 1.150      &&&&  -0.3815  &&&&1.406  \\
YS-III    &&&&              &&&& 1     &&&&  -0.1323   &&&& 1.130      &&&&  -0.3815  &&&&1.406  \\
YS-IV     &&&&              &&&& 1     &&&&  -0.1323   &&&& 1.130      &&&&  -0.3628  &&&&1.406  \\
\\
MT-I/III  &&&&   Yukawa     &&&& 1     &&&&  7.39     &&&& 3.110      &&&& 7.39     &&&& 3.110 \\
          &&&&              &&&& 2     &&&&  -2.64    &&&& -1.555     &&&& -3.22    &&&& -1.555 \\
\\
$P_{5.5}GL$ &&&& Separable  &&&& 1     &&&& 0.13230   &&&& 1.130      &&&& -0.18752  &&&& 1.2766\\
            &&&&            &&&& 2     &&&&           &&&&            &&&& -0.18752  &&&& 1.7610\\
\hline \hline
\end{tabular}
\end{center}
\end{table}

Our results will be compared to several techniques: the VAR \cite{Fantoni-NCA69}
and HH \cite{Ballot-ZPA302}-\cite{Viviani-PRC71} methods, several
types of approximations for the subsystem kernels of the four-body problem by
operators of finite rank (SKFR) \cite{Sofianos-PRC26}-\cite{Tjon-PLB63}, the
integrodifferential equation approaches SIDE \cite{Oehm-PRC42} and IDEA
\cite{Ellerkman-PRC53}, the CRC \cite{Yakovlev-PAN60},
the DFY \cite{Schellingerhout-PRC46,Merkuriev-NPA431}, the FY (PW)
\cite{Kamada-NPA548}, and last, but not least, 2DI \cite{Gibson-PRC15}.
Our results for the triton and $\alpha$-particle binding energies are shown
in tables \ref{table.S3}-\ref{table.P5.5} in comparison to the results of
other techniques. Table \ref{table.S3} collects the binding energies
for the S$_{3}$ potential, Table \ref{table.Y's} for the YS type potentials,
Table \ref{table.MT} for the MT-I/III potential, and Table \ref{table.P5.5} for the
$P_{5.5}GL$ potential.

As shown in Table \ref{table.S3}, our result for the $\alpha$-particle binding
energy for the spin-dependent (spin-averaged) S$_{3}$ potential with value
$-28.8 \,(-25.7) \, $ MeV is in good agreement with results of HHE, SIDE, DFY
techniques and especially with FY result in PW decomposition. Also, our result for
the triton binding energy with values $-8.20$ and $-6.41 \, $ MeV, corresponding to
spin-dependent and -averaged versions of this potential, are in excellent agreement
with FY results in PW decomposition. It should be pointed that the results with
spin-averaged version of the potentials differ from previous results where
the original version of the potentials was used. The difference between obtained
results of original and averaged versions of the potentials is to be expected and it
is quite natural.

\begin{table}
\caption{\label{table.S3} Triton and $\alpha$-particle binding energies for S$_{3}$
potential in MeV.}
\begin{center}
\begin{tabular}{lllllllllllllllllllll} \hline \hline
Method &&&&&&&&&  $E_t$  &&&&&&&&&  $E_{\alpha}$  \\
\hline
VAR \cite{Fantoni-NCA69}          &&&&&&&&&       &&&&&&&&& -26.47 \\
HHE \cite{Ballot-FBSS1}           &&&&&&&&&       &&&&&&&&& -26.01 \\
SIDE \cite{Oehm-PRC42}            &&&&&&&&& -8.20 &&&&&&&&& -27.93  \\
CRC \cite{Yakovlev-PAN60}         &&&&&&&&&       &&&&&&&&& -28.74 \\
DFY \cite{Schellingerhout-PRC46}  &&&&&&&&&       &&&&&&&&& -28.79 \\
FY(PW) \cite{Kamada-NPA548}       &&&&&&&&& -8.20 &&&&&&&&& -28.80   \\
FY(3D)                            &&&&&&&&& -8.20 &&&&&&&&&  -28.8 \\
\hline
SIDE$^{av}$ \cite{Oehm-PRC42}            &&&&&&&&&       &&&&&&&&& -25.38  \\
DFY$^{av}$ \cite{Schellingerhout-PRC46}  &&&&&&&&&       &&&&&&&&& -25.50      \\
HHE$^{av}$ \cite{Ballot-ZPA302}          &&&&&&&&&       &&&&&&&&& -25.97  \\
DFY$^{av}$ \cite{Merkuriev-NPA431}       &&&&&&&&&       &&&&&&&&& -25.68      \\
FY(PW)$^{av}$ \cite{Kamada-NPA548}       &&&&&&&&& -6.41 &&&&&&&&& -25.69  \\
FY(3D)$^{av}$                            &&&&&&&&& -6.41 &&&&&&&&& -25.7 \\
\hline
Exp. &&&&&&&&& -8.48 &&&&&&&&& -28.30
\\ \hline \hline
\end{tabular}
\end{center}
\end{table}

The calculated triton and $\alpha$-particle binding energies for separable, spin-dependent
Yamaguchi type potentials with different methods are listed in Table \ref{table.Y's}.
Our results for the $\alpha$-particle (triton) binding energy for YS I, II, III and IV
with values $-45.9\, (-11.05)$, $-44.4\, (-10.70)$, $-42.4\, (-10.13)$, $-37.8\, (-8.47) \, $
MeV, are in excellent agreement with the 2DI results.

\begin{table*}
\caption{\label{table.Y's} $\alpha$-particle binding energy for YS-type
potentials in MeV. The numbers in parenthesis are corresponding to triton binding
energies. }
\begin{center}
\begin{tabular}{llllllllllllllllllllll}  \hline \hline
Method  &&&& YS-I &&&& YS-II &&&& YS-III &&&& YS-IV \\ \hline
FY(PW) \cite{Kamada-NPA548}  &&&&  -45.87 (-11.05) \\
SKFR \cite{Narodetsky-PLB46} &&&& -45.73  \\
SKFR \cite{Fonseca-PRC30}    &&&& -45.59  \\
SKFR \cite{Fonseca-FBS1}     &&&& -45.32  \\
2DI  \cite{Gibson-PRC15}     &&&& -45.7 (-11.05) &&&&  -44.2 (-10.71) &&&&  -42.3 (-10.13) &&&& -37.7 (-8.48) \\
FY(3D)    &&&& -45.9 (-11.05)  &&&&  -44.4 (-10.70)  &&&& -42.4 (-10.13)  &&&& -37.8 (-8.47)  \\
\hline Exp. &&&& &&&& -28.30 (-8.48)  \\ \hline \hline
\end{tabular}
\end{center}
\end{table*}

As demonstrated in Table \ref{table.MT}, the calculation of the $\alpha$-particle binding energy
by using the spin-dependent and spin-averaged version of MT-I/III potential in the FY(PW) scheme
converges to values of $-30.29 \, $ and $-28.83 \, $ MeV, while the triton binding energy converges
to values $-8.54$ and $-7.55 \, $ MeV, correspondingly. As shown in this table our calculations for
spin-dependent version of this potential yields the values $-8.54$ and $-30.3\, $ MeV for triton and
$\alpha$-particle binding energies correspondingly, which are in good agreement with the FY (PW) results.
Also, our results for the triton and $\alpha$-particle binding energies with the spin-averaged version
of this potential with values $-7.57 \, $ and $-28.8 \, $ MeV are also in excellent agreement with
the corresponding FY (PW) results.

\begin{table*}
\caption{\label{table.MT} Triton and $\alpha$-particle binding energies for
Malfliet-Tjon I/III potential in MeV.}
\begin{center}
\begin{tabular}{lllllllllllllllllll}  \hline \hline
Method  &&&&&&& $E_t$  &&&&&&& $E_{\alpha}$    \\
\hline
SKFR \cite{Tjon-PLB63}               &&&&&&&           &&&&&&& -29.6 \\
SKFR \cite{Sofianos-PRC26}           &&&&&&&           &&&&&&& -30.36   \\
SIDE \cite{Oehm-PRC42}               &&&&&&&  -8.54    &&&&&&& -29.74   \\
DFY \cite{Schellingerhout-PRC46}     &&&&&&&  -8.54   &&&&&&& -30.31   \\
IDEA \cite{Ellerkman-PRC53}          &&&&&&& -8.86     &&&&&&&  -30.20   \\
HH \cite{Viviani-PRC71}              &&&&&&&           &&&&&&&  -30.33   \\
EIHH \cite{Barnea-PRC61}                        &&&&&&& -8.72    &&&&&&&  -30.71   \\
DFY(PW) \cite{Schellingerhout-PRC46}                      &&&&&&&           &&&&&&&  -30.312   \\
FY(PW) \cite{Kamada-NPA548}          &&&&&&& -8.54     &&&&&&& -30.29 \\
FY(3D)           &&&&&&& -8.54     &&&&&&& -30.3  \\
\hline
FY(PW)$^{av}$ \cite{Kamada-NPA548}   &&&&&&& -7.55     &&&&&&& -28.83 \\
FY(3D)$^{av}$                        &&&&&&& -7.55     &&&&&&& -28.8
\\ \hline
 Exp.                                &&&&&&& -8.48     &&&&&&& -28.30  \\ \hline
 \hline
\end{tabular}
\end{center}
\end{table*}

In Table \ref{table.P5.5}, we present the triton and $\alpha$-particle binding energies for the
$P_{5.5}GL$ potential calculated with the SKFR and FY methods. Our results for triton and $\alpha$-particle
binding energies with values $-8.04$ and $-28.9$ MeV are in excellent agreement with the corresponding PW
results. In the next section, we present our results for binding energies with the inclusion of 3NFs.

\begin{table*}
\caption{\label{table.P5.5} Triton binding energy for
$P_{5.5}GL$ potential in MeV. The numbers in parenthesis are $\alpha$-particle binding
energies.}
\begin{center}
\begin{tabular}{lllllllllllllllllll}  \hline \hline
Method  &&&&&&& $E_t$   \\
\hline
SKFR \cite{Fonseca-PRC30}     &&&&&&& -29.10 \\
FY(PW) \cite{Kamada-NPA548}   &&&&&&&   -28.87 (-8.04) \\
FY(3D)    &&&&&&&  -28.9 (-8.04) &&&&&&&   \\
\hline Exp. &&&&&&& -28.30 (-8.48)
\\ \hline \hline
\end{tabular}
\end{center}
\end{table*}

\subsection{Results for $NN$ with $3N$ Potential Models}

 In our calculations with a 3NF, we use a model of the 3NF which is based on multi-meson exchanges.
 We study two different types of 3NFs, a purely attractive and a superposition of attractive and repulsive,
 which are named MT3-I and MT3-II respectively, Ref. \cite{Liu-FBS33}. The parameters of these 3NFs are
 chosen so that the correction due to these 3NFs to the triton binding energy calculated with the modified
 Malfliet-Tjon (MT2-II) NN potential is small, and they lead to binding energies near to the experimental
 triton binding energy.

As shown in Table \ref{table.3BF}, our results for the $\alpha$-particle (triton) binding energies with the
addition of the MT3-I and MT3-II 3NFs, while the averaged version of MT-I/III is used as the NN potential,
are $-35.7\,(-8.68)$ and $-34.5\, (-8.45)$ MeV, respectively. Unfortunately we could not compare these
results for binding energies with other calculations, but we have listed our recent results with different
combination of MT-V $NN$ potential and mentioned 3N potential models, i.e. MT3-I and MT3-II,
\cite{Hadizadeh-EPJA36}. As one can see from the comparison of our results with and without 3NFs (while
MT-I/III$^{ave}$ is used as NN potential model) with the previously calculated binding energies (while MT-V
is used as NN potential model) the MT-I/III$^{ave}$ NN potential model provide more reasonable results in
comparison to MT-V for triton and $\alpha$-particle binding energies.

% Three- and four-body binding energies
\begin{table}
\caption{\label{table.3BF} Triton and $\alpha$-particle binding energies with
and without 3NFs in MeV. }
\begin{center}
\begin{tabular}{lllllllllllllllllllllllllll}
\hline \hline
Potential  &&&&&&& $E_t$  &&&&&&& $E_{\alpha}$  \\
\hline
MT-I/III$^{ave}$            &&&&&&&  -7.55 &&&&&&& -28.8 \\
MT-I/III$^{ave}$+MT3-I      &&&&&&&  -8.68 &&&&&&& -35.7 \\
MT-I/III$^{ave}$+MT3-II     &&&&&&&  -8.45 &&&&&&& -34.5 \\
\hline
MT-V \cite{Hadizadeh-FBS40}           &&&&&&&  -7.74 &&&&&&& -31.3 \\
MT-V+MT3-I \cite{Hadizadeh-EPJA36}      &&&&&&&  -8.92 &&&&&&& -38.8 \\
MT-V+MT3-II  \cite{Hadizadeh-EPJA36}   &&&&&&&  -8.70 &&&&&&& -37.5 \\
\hline Exp. &&&&&&& -8.48 &&&&&&& -28.30
\\ \hline
\hline
\end{tabular}
\end{center}
\end{table}

All these numbers are not meant to provide insight into the physics of three and four interacting
nucleons, but have the purpose to demonstrate the high accuracy of numerical results that
one can obtain by considering the present non-PW approach, in face of other existent methods.
The advantages of the method relies in a simplified and straightforward formalism, which is
appropriate to treat typical nuclear forces consisting of attractive and repulsive (short range)
parts. The results presented indicate that the 3D approach leads to numerical results with the same
accuracy of PW-based methods, whereas it leads to integral equations with much less analytical and
algebraic complexity in comparison to corresponding equations formulated in PW-based methods.
In a 3D case, there are only a finite number of coupled three-dimensional integral equations to be
solved; whereas, in the PW case, after truncation, one has a finite number of coupled equations with
kernels containing relatively complicated geometrical expressions.

\section{Summary and outlook}\label{sec: Summary}

In summary, in the present paper we solve the FY three-dimensional integral equations for spin-dependent and spin-averaged NN potential models, i.e., S$_{3}$, MT I/III, YS-type and
$P_{5.5}GL$ and the scalar two-meson exchange three-body interaction. These potentials provide
reasonable results for binding energies in comparison to the potential models that have been
used in previous works. Our results for these potential models are in good agreement with the corresponding previous values when considering VAR, HHE, SKFR, SIDE and DFY techniques. In
particular, they are matched with PW calculations in the FY scheme.

This non-PW approach, by directly working with momentum vector variables, is being revealed
as an efficient good alternative to other methods to treat three- and four-nucleon
bound-state calculations. Recently, following this approach, the coupled FY equations have been formulated with and
without 3NFs, as a function of vector Jacobi momenta, where the formalism is given in terms
of the magnitudes of the momenta and the angles between them. It has been demonstrated
that the three-dimensional FY integral equations can be handled in a straightforward and
numerically reliable fashion. In comparison to commonly used angular momentum decompositions,
this direct approach leads to a finite number of coupled equations with kernels containing
very simplified expressions.

It should be clear that this approach is more efficient for scattering problems, especially
in the energy regions where the PW-based calculations have slow convergence. The formulation
of 3N scattering and $^3$H photodisintegration in a realistic 3D approach has been done successfully \cite{Bayegan-EPJWC3}-\cite{Harzchi-EPJA46} and the calculation is underway. Molecular, atomic, and
nuclear or subnuclear physics are but a few examples of various fields of physics where quantum
mechanical few-body problems play an important role. Since the 3D approach is general,
it can be applied to any system from molecules to elementary particles.
Another valuable application of this non-PW approach, are the few-body atomic bound states with realistic potentials.

We should also mention a renormalization group approach that our group has considered
when solving integral equations for the nucleon-nucleon (NN) interaction \cite{npa99}.
In leading order, by using the one-pion-exchange potential (OPEP) plus a Dirac-delta function,
is considered a non-perturbative renormalization procedure, relying on a subtracted kernel
where a scaling parameter is introduced. The role of the scaling parameter is similar to the
cut-off momentum parameter but with a big advantage in view of its flexibility.
Since the approach is renormalization group invariant, one can arbitrarily move the reference
scale without affecting the relevant physical results. An extension of this approach is being submitted for publication \cite{timoteo-2010}, where a recursive subtraction procedure is applied
to the scattering matrix solution with next-leading-order (NLO) and next-to-next-leading-order (NNLO) two-pion exchange interactions. Also, we are considering the application of the present
3D approach for the NN interaction in the renormalization group scheme that was used
in Ref. \cite{npa99}.

\section*{acknowledgments}
M. R. Hadizadeh and L. Tomio would like to thank the Brazilian agencies FAPESP and CNPq for partial
support. S. Bayegan acknowledges the support of the center of excellence of University of Tehran.


\begin{thebibliography}{0}

\bibitem{Schellingerhout-PRC46} N. W. Schellingerhout, J. J. Schut, and L. P. Kok, Phys. Rev. C {\bf 46}, 1192 (1992).

\bibitem{Lazauskas-PRC70} R. Lazauskas and J. Carbonell, Phys. Rev. C {\bf 70}, 044002  (2004).

\bibitem{Kamada-NPA548} H. Kamada and W. Gl\"{o}ckle, Nucl. Phys. A {\bf 548}, 205 (1992).

\bibitem{Glockle-NPA560} W. Gl\"{o}ckle and H. Kamada, Nucl. Phys. A {\bf 560}, 541 (1993).

\bibitem{Nogga-PRL85} A. Nogga, H. Kamada, and W. Gl\"{o}ckle, Phys. Rev. Lett. {\bf 85}, 944 (2000).

\bibitem{Kamada-PRC64} H. Kamada et al., Phys. Rev. C {\bf 64}, 044001 (2001).

\bibitem{Nogga-PRC65} A. Nogga, H. Kamada, W. Gl\"{o}ckle, and B. R. Barrett, Phys. Rev. C {\bf 65}, 054003 (2002).

\bibitem{Epelbaum-EPJA15} E. Epelbaum et al., Eur. Phys. J. A {\bf 15 }, 543 (2002).

\bibitem{Epelbaum-PRC70} E. Epelbaum et al., Phys. Rev. C {\bf 70}, 024003 (2004).

\bibitem{Bayegan-PRC77} S. Bayegan, M. R. Hadizadeh, and W. Gl\"{o}ckle, Prog. Theor. Phys. {\bf 120}, 887 (2008); S. Bayegan, M. R. Hadizadeh, and M. Harzchi, Phys. Rev. C {\bf 77}, 064005 (2008); M. R. Hadizadeh and S. Bayegan, Mod. Phys. Lett. A {\bf 24}, 816 (2009).

\bibitem{Elster-FBS27} Ch. Elster, W. Schadow, A. Nogga, and W. Gl\"{o}ckle, Few Body Syst. {\bf 27}, 83 (1999).

\bibitem{Liu-FBS33} H. Liu, Ch. Elster, and W. Gl\"{o}ckle, Few Body Syst. {\bf 33}, 241 (2003).

\bibitem{Hadizadeh-FBS40} M. R. Hadizadeh and S. Bayegan, Few Body Syst. {\bf 40}, 171 (2007).

\bibitem{Hadizadeh-EPJA36} M. R. Hadizadeh and S. Bayegan, Eur. Phys. J. A {\bf 36}, 201 (2008).

\bibitem{Liu-PRC72} H. Liu, Ch. Elster, and W. Gl\"{o}ckle, Phys. Rev. C {\bf 72}, 054003 (2005); T. Lin, Ch. Elster, W.N. Polyzou, and W. Gl\"{o}ckle, Phys. Lett. B {\bf 660}, 345 (2008); Ch. Elster, T. Lin, W. Gl\"{o}ckle, and S. Jeschonnek, Phys. Rev. C {\bf 78}, 034002 (2008); Ch. Elster, W. Gl\"{o}ckle, and H. Witala, Few Body Syst. {\bf 45}, 1 (2009).

\bibitem{Witala-EPJA41} H. Witala , R. Skibinski, J. Golak, and W. Gl\"{o}ckle, Eur. Phys. J. A. {\bf 41}, 369 (2009);  H. Witala, R. Skibinski, J. Golak, and W. Gl\"{o}ckle, Eur. Phys. J. A. {\bf 41}, 385 (2009).


\bibitem{Fachruddin-PhD} I. Fachruddin, Ph.D. thesis, Ruhr-Universit\"{a}t Bochum, (2003).


\bibitem{Gloeckle-FBS47} W. Gl\"{o}ckle, Ch. Elster, J. Golak, R. Skibinski, H. Witala, and H. Kamada, Few Body Syst.{\bf 47}, 25 (2010);
W. Gl\"{o}ckle, I. Fachruddin, Ch. Elster, J. Golak, R. Skibinski, and H. Witala, Eur. Phys. J. A {\bf 43}, 339 (2010).

\bibitem{Afnan-PR175} I. R. Afnan and Y. C. Tnag, Phys. Rev. {\bf 175}, 1337 (1968).

\bibitem{Gibson-PRC15} B. F. Gibson and D. R. Lehman, Phys. Rev. C {\bf 15}, 2257 (1977).

\bibitem{Malfliet-NPA127} R. A. Malfliet and J. A. Tjon, Nucl. Phys. A {\bf 127}, 161 (1969).

\bibitem{Gibson-PRC18} B. F. Gibson and D. R. Lehman, Phys. Rev. C {\bf 18}, 1042 (1978).

\bibitem{Fantoni-NCA69} S. Fantoni, L. Panattoni and S. Rosati, Nuovo Cimento A {\bf 69}, 80 (1970).

\bibitem{Ballot-ZPA302} J. A. Ballot, Z. Phys. A {\bf 302}, 347 (1981).

\bibitem{Ballot-FBSS1} J. A. Ballot, Few Body Syst. Suppl. {\bf 1}, 140 (1986).

\bibitem{Barnea-PRC61} N. Barnea, W. Leidemann, and G. Orlandini, Phys. Rev. C {\bf 61}, 054001 (2000).

\bibitem{Viviani-PRC71} M. Viviani, A. Kievsky and S. Rosati, Phys. Rev. C {\bf 71}, 024006 (2005).

\bibitem{Sofianos-PRC26} S. A. Sofianos, H. Fiedeldey, H. Haberzettel, and W. Sandhas, Phys. Rev. C {\bf 26}, 228 (1982).

\bibitem{Fonseca-PRC30} A. C. Fonseca, Phys. Rev. C {\bf 30}, 35 (1984).

\bibitem{Fonseca-FBS1} A. C. Fonseca, Few Body Syst. {\bf 1}, 69 (1986).

\bibitem{Narodetsky-PLB46} I. M. Narodetsky, E. S. Galpern and V. N. Lyakhovitsky, Phys. Lett. B {\bf 46}, 51 (1973);
I. M. Narodetsky, Nucl. Phys. A {\bf 221}, 191 (1974).

\bibitem{Tjon-PLB63} J. A. Tjon, Phys. Lett. B {\bf 63}, 391 (1976).

\bibitem{Oehm-PRC42} W. Oehm, S. A. Sofianos, H. Fiedeldey, and M. Fabre de la Ripelle, Phys. Rev. C {\bf 42}, 2322 (1990).

\bibitem{Ellerkman-PRC53} G. G. Ellerkman, W. Sandhas, S. A. Sofianos, and H. Fiedeldey, Phys. Rev. C {\bf 53}, 2638 (1996).

\bibitem{Yakovlev-PAN60} S. L. Yakovlev and I. N. Filikhin, Phys. Atom. Nucl. {\bf 60}, 1794 (1997).

\bibitem{Merkuriev-NPA431} S. B. Merkuriev, S. L. Yakovlev, and C. Gignoux, Nucl. Phys. A {\bf 431}, 125 (1984).

\bibitem{Bayegan-EPJWC3} S. Bayegan, M. A. Shalchi, and M. R. Hadizadeh, EPJ Web of Conferences {\bf 3}, 04008 (2010).

\bibitem{Harzchi-EPJA46} M. Harzchi and S. Bayegan, Eur. Phys. J. A {\bf 46}, 271 (2010).

\bibitem{npa99} T. Frederico, V. S. Tim\'oteo and L. Tomio, Nucl. Phys. A {\bf 653}, 209 (1999);
T. Frederico, A. Delfino and L. Tomio, Phys. Lett. B {\bf 481}, 143 (2000);
V.S. Tim\'oteo, T. Frederico, A. Delfino, and L. Tomio, Phys. Lett. B {\bf 621}, 109 (2005).

\bibitem{timoteo-2010} V. S. Tim\'oteo, T. Frederico, A. Delfino, and L. Tomio,
arXiv:1006.1942.


\end{thebibliography}
\end{document}